\documentclass[aps,prx,reprint]{revtex4-1}

\usepackage{graphicx}
\usepackage{mathtools,amsthm,amssymb,braket}

\usepackage[utf8]{inputenc}
\usepackage[german,  danish, english]{babel}

\usepackage[pdfauthor={V. O. Shkolnikov, and Guido Burkard},
pdftitle={Theory of generalized Lambda systems},
pdfsubject={},colorlinks=true,linkcolor=blue,citecolor=magenta]{hyperref}

\usepackage{bbm}

\begin{document}

%%%%%%%%%%%%%%%%%%%%%%%%%%%%%%%%%%%%%%%%%%%%%%%%%%%%%%%%%%%%%%%

\title{Effective Hamiltonian theory of the geometric evolution of quantum systems}

\author{V. O. Shkolnikov}
\author{Guido Burkard}
\affiliation{Department of Physics, University of Konstanz, D-78457 Konstanz, Germany}

%\date{\today}

%%%%%%%%%%%%%%%%%%%%%%%%%%%%%%%%%%%%%%%%%%%%%%%%%%%%%%%%%%%%%%%
%%% Abstract
%%%%%%%%%%%%%%%%%%%%%%%%%%%%%%%%%%%%%%%%%%%%%%%%%%%%%%%%%%%%%%%

\begin{abstract}
	In this work we present an effective Hamiltonian description of the quantum dynamics of a generalized Lambda system undergoing adiabatic evolution. We assume the system to be initialized in the dark subspace and show that its holonomic evolution can be viewed as a conventional Hamiltonian dynamics in an appropriately chosen extended Hilbert space. In contrast to the existing approaches, our method does not require the calculation of the non-Abelian Berry connection and can be applied without any parametrization of the dark subspace, which becomes a challenging problem with increasing system size.
\end{abstract}

\maketitle
%%%%%%%%%%%%%%%%%%%%%%%%%%%%%%%%%%%%%%%%%%%%%%%%%%%%%%%%%%%%%%%
%%% Bibliography
%%%%%%%%%%%%%%%%%%%%%%%%%%%%%%%%%%%%%%%%%%%%%%%%%%%%%%%%%%%%%%%
\textit{Introduction.}$\--$Quantum information science is an active and developing field of study which has motivated an intense search for physical systems that can be used as quantum processors.  Whatever system is eventually going to be used, one must be able to efficiently manipulate the state of the quantum device with high gate fidelity in order to either perform sufficiently long quantum computations without error correction  \cite{Preskill} or to allow for fault-tolerant quantum operation \cite{Campbell}.  Most proposed and implemented quantum processors use dynamical protocols to manipulate the quantum state of the device by controlling a non-zero Hamiltonian $H(t)$ acting on the quantum register directly to generate the time evolution $U={\rm T}\exp(-i\int H(t)dt)$.  Alternatively, geometric phases \cite{Berry} and their non-Abelian generalizations arising after a cyclic adiabatic evolution of the system can be used to realize universal quantum gates \cite{Zanardi_Rosetti, Sjoeqvist3}.  In this case the system is initialized in the dark subspace of its Hamiltonian and due to the adiabatic theorem remains there as the Hamiltonian is slowly changed in time. This method can provide an intrinsic tolerance against certain types of noise \cite{Shi-Liang Zhu_Paolo Zanardi,Childs_Edward Farhi_John Preskill} and was realized experimentally in NMR systems \cite{NMR_Jones_Vedral_Ekert_Giuseppe Castagnoli}. Later, proposals based on tripod systems \cite{Duan_Cirac_Zoller,Kis_Renzoni,Moeller_Madsen_Moelmer} were experimentally applied to realize single qubit rotations in trapped ions \cite{Trapped_ions_Toyoda_Uchida_Noguchi_Haze_Urabe}. There are other proposals to realize geometric gates in systems of superconducting qubits \cite{Kamleitner_Solinas_Mueller_Shnirman_Moettoenen}. Geometric gates can also be constructed using non-adiabatic evolution \cite{Sjoeqvist, Sjoeqvist2}. Relaxing the adiabaticity condition makes it simpler to perform gates, and non-adiabatic geometric gates have indeed been successfully realized with superconducting qubits \cite{Abdumalikov}, NMR systems \cite{Feng} and with the electron spin of nitrogen-vacancy centers  \cite{Balasubramanian,Zu,Zhou}. Here, we will restrict ourselves to the adiabatic case.

In this paper we consider the time evolution of a quantum system initialized in the instantaneous dark subspace of its time-dependent Hamiltonian, as required for adiabatic geometric quantum computation. A conventional way to describe the dynamics of such systems would be to find a basis in the instantaneous dark subspace, compute the non-Abelian Berry connection using this basis \cite{Shapere_Wilczek,Wilczek_Zee}, and subsequently evaluate the path-ordered exponential of the line integral of the obtained Berry connection along the path in the Hamiltonian parameter space. We show that it is possible to describe the evolution of this system without explicitly calculating the Berry connection, but by introducing an effective Hamiltonian and then solving the Schr\"{o}dinger equation instead. The Hamiltonian used for this procedure acts in a Hilbert space large enough to contain the instantaneous dark subspace at any moment in time. It may coincide with the full Hilbert space of the system, but can also be smaller if the dark subspace never involves some of the system's levels. Our approach suggests that instead of computing a basis in the instantaneous dark subspace of the time dependent Hamiltonian one can identify its bright states and use them to construct an effective Hamiltonian that contains all the information about the adiabatic evolution of the dark subspace. That means that a complicated procedure of finding the orthonormal basis in the possibly very large dark space of the system can be avoided, which makes the numerical description of the system dynamics much less demanding.

\begin{figure}[t]
	\includegraphics[width=0.45\textwidth]{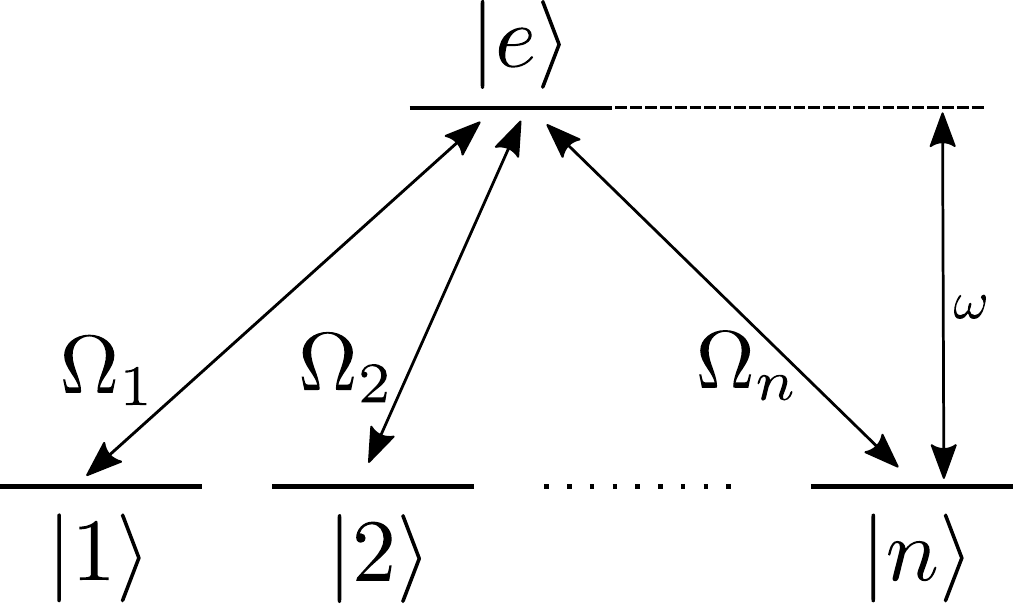}
	\caption{Energy level diagram of the generalized Lambda-system. The lower $n$ states $\ket{i}$ ($i=1,...,n$) are coupled to the excited state $\ket{e}$. The system has an instantaneous $(n-1)-$dimensional dark subspace, in which it remains due to the adiabatic theorem.
		\label{fig:Lambda_system}}
\end{figure}

\textit{System description.}$\--$We first consider a generalized Lamda system with $n+1$ levels, for which the first $n$ levels, forming a Hilbert space $\mathcal{H}$ are separated from the remaining level with the energy $w$ and are resonantly coupled to it by oscillating fields $\Omega_ie^{iwt}$ 
(Fig.~\ref{fig:Lambda_system}). From here on, we use units in which $\hbar=1$. The detailed description of the adiabatic evolution of this system, obtained with the formalism of the non-Abelian Berry connection, is well known in the literature \cite{Recati_Calarco_Zanardi_Cirac_Zoller}. The Hamiltonian of such a system in the rotating frame is
\begin{equation}
\hat{H}=\sum_{i=1}^{n}
\left(
\Omega_i\ket{i}\bra{e}+\Omega_i^*\ket{e}\bra{i}
\right)
\label{Lambda_Hamiltonian},
\end{equation}
where $\Omega_i$ is the complex coupling amplitude (Rabi frequency) of the $i^{th}$ level $\ket{i}$ to the excited state $\ket{e}$ (Fig.~\ref{fig:Lambda_system}). It should be noted here that the form of the Hamiltonian (\ref{Lambda_Hamiltonian}) does not require the ground states to be degenerate. It suffices that each ground state level $\ket{i}$ is coupled to $\ket{e}$ resonantly. This is the reason why the rotating frame Hamiltonian (\ref{Lambda_Hamiltonian}) can arise in many different multilevel systems, the requirement being the absence of parasitic coupling between the ground states.

We introduce the mean Rabi frequency $\Omega=\sqrt{\sum_{i=1}^{n}\Omega_i^2}$ and parametrize the coupling coefficients as $\Omega_i/\Omega=r_ie^{i\phi_i}$, so that the Hamiltonian (\ref{Lambda_Hamiltonian}) can be rewritten as
\begin{equation}
\hat{H}=\Omega\sum_{i=1}^{n}r_i\left(e^{i\phi_i}\ket{i}\bra{e}+e^{-i\phi_i}\ket{e}\bra{i}\right).
\label{Hamiltonian}
\end{equation}
Here $r_i$ are positive numbers obeying the property $\sum_{i=1}^{n}r_i^2=1$. By introducing the normalized bright state, 
\begin{equation}
\ket{B}=\sum_{i=1}^{n}r_ie^{i\phi_i}\ket{i},
\label{bright_state_definition}
\end{equation}
the Hamiltonian of the system can be rewritten as
\begin{equation}
\hat{H}=\Omega\left(\ket{B}\bra{e}+\ket{e}\bra{B}\right).
\label{one_bright_state_hamiltonian}
\end{equation}
The values of the $n$ amplitudes and $n$ phases of the excitation fields determine the bright state, thus defining a Hamiltonian that acts trivially on the orthogonal complement of the space spanned by $\ket{B}$ in the Hilbert space $\mathcal{H}$. These $n-1$ states form the so-called dark subspace, such that for any state $\psi$ in this subspace $\hat{H}\psi=0$. If the bright state is specified, one can uniquely define the dark subspace as its orthogonal complement. The global phase of the bright state is not important for the identification of the dark subspace; if one also takes into account the normalization condition for the $r_i$, one concludes that one has $2n-2$ independent parameters that define the dark subspace of the Hamiltonian $\hat{H}$. Considering a time-dependent excitation with $r_i(t)$ and $\phi_i(t)$, the evolution of the system is governed by a time dependent Hamiltonian $\hat{H}(t)$, whose dark subspace is now time dependent and describes a path in the $(2n-2)-$parametric space. In the adiabatic regime, where the parameters are changed slowly with respect to $1/\Omega$, it is known that if the system starts in the dark subspace, it will remain there during the evolution \cite{Born_Fock}. The Initialization of the system in the dark subspace is by itself an interesting issue and strongly depends on the system. In the case when the state $\ket{e}$ has a short lifetime a way to initialize the system would be to use coherent population trapping (CPT) \cite{Fleischhauer}. If we do not restrict the Hamiltonian to contain the couplings to the state $\ket{e}$ only, but allow couplings between the ground states as well, we can construct a Hamiltonian that contains only one dark state, for example $\ket{1}$. If the system does not start in the state $\ket{1}$, it will be pumped with such a Hamiltonian to $\ket{e}$, from which it may either decay to $\ket{1}$ or to some other ground state and then the process will repeat. After many cycles the system will be trapped in the state $\ket{1}$. Then one can switch off the couplings between the ground states and return to the case of the Hamiltonian (\ref{one_bright_state_hamiltonian}), assuming the system is initialized in its dark subspace.  The standard way to describe the adiabatic evolution of this subspace would be to write down the basis vectors in the dark subspace, that depend on the $2n-2$ independent parameters and then calculate the Berry connection. The path ordered exponential of the line integral of the Berry connection will then define the evolution operator of the system. In what follows we will present a different formalism to analyze the evolution of the system based on the construction of an effective Hamiltonian in the whole Hilbert space $\mathcal{H}$ and discuss the purposes to which it could be applied.

 \textit{Effective Hamiltonian.}$\--$Let $\ket{\psi_1(t)}$, $\ket{\psi_2(t)}$,......, $\ket{\psi_{n-1}(t)}$ be orthonormal basis in the dark subspace of the system at time $t$. We assume the system to be initialized in an instantaneous dark state of its Hamiltonian and to subsequently evolve in the adiabatic regime, so at any time moment $t$ the state of the system is $\ket{\psi_s}=\sum_{i=1}^{n-1}c_i(t)\ket{\psi_i(t)}$. 
Let us now concentrate on how this general state evolves due to the Hamiltonian $\hat{H}(t)$ during an infinitesimal time interval $dt$. In Appendix~\ref{appA} we show that
\begin{equation}
\ket{\psi_s}\rightarrow \hat{U}\ket{\psi_s},
\end{equation}
where the unitary operator $\hat{U}$ acts in the whole Hilbert space $\mathcal{H}$ and is given by
\begin{equation}
\begin{split}
\hat{U}=\hat{1}+\left[\dot{\ket{B}}\bra{B}-\ket{B}\bra{\dot{B}}\right]dt,
\label{Unitary_lambda_system}
\end{split}
\end{equation}
where $\hat{1}$ is the projection operator on the Hilbert space $\mathcal{H}$, acting as identity in this space. Since the operator $\hat{U}$ generates the correct evolution of the states in the dark subspace of the system, we can view the evolution of the state in the dark subspace in the adiabatic regime as if a time dependent effective Hamiltonian
\begin{equation}
\hat{H}_{\text{eff}}=i\left(\dot{\ket{B}}\bra{B}-\ket{B}\bra{\dot{B}}\right)
\label{effective_hamiltonian}
\end{equation}
was acting in the Hilbert space $\mathcal{H}$.

In terms of the laser coupling coefficients $r_i,\phi_i$ $(i=1,....,n)$, using equation (\ref{bright_state_definition}), the effective Hamiltonian (\ref{effective_hamiltonian}) can be rewritten in the original basis $\ket{i}$ $(i=1,....,n)$ as
\begin{equation}
\hat{H}_{\text{eff}}=\sum_{i,j=1}^{n}r_ir_j\left[-(\dot{\phi}_i+\dot{\phi}_j)+i\frac{d}{dt}\ln\left(\frac{r_i}{r_j}\right)\right]e^{i(\phi_i-\phi_j)}\ket{i}\bra{j}.
\label{laser_couplings_effective Hamiltonian}
\end{equation} 
In Appendix~\ref{appC} we show that this Hamiltonian can describe the same universal set of gates as the non-Abelian Berry connection, demonstrating that the two approaches are indeed equivalent.

\textit{Generalizations.}$\--$Let us now consider a more general Hamiltonian of the form
\begin{equation}
\hat{H}=\sum_{i,j=1}^{k}\left(g_{ij}\ket{B_i}\Bra{B_j}+g_{ij}^*\ket{B_j}\Bra{B_i}\right),
\label{more_general_Hamiltonian}
\end{equation}
acting in some Hilbert space $\mathcal{H}$ of dimension $n$, where $\ket{B_i}$ are time dependent states in this Hilbert space, forming an orthonormal set of vectors at any instant in time. Note that the Hamiltonian (\ref{one_bright_state_hamiltonian}) is a special case of (\ref{more_general_Hamiltonian}), with two bright states, one of them being constant. We point out that the Hamiltonian (\ref{more_general_Hamiltonian}) can always be brought into a diagonal form with appropriately chosen bright states, but for our purposes it is not necessary to assume this.

From now on we will assume that the instantaneous eigenvalues of the Hamiltonian (\ref{more_general_Hamiltonian}) in the subspace spanned by the vectors $\ket{B_{i}}$ are nonzero and that the adiabatic condition with respect to these eigenvalues is fulfilled. Thus, if the system starts in the instantaneous dark subspace of this Hamiltonian, it never leaves it, in accordance with the adiabatic theorem. 

In full analogy to the case of equation (\ref{Unitary_lambda_system}), in Appendix~\ref{appB} we show that one can build the unitary transformation acting in the Hilbert space $\mathcal{H}$ and describing correctly the transformation of the dark subspace of Hamiltonian (\ref{more_general_Hamiltonian}) during the infinitesimal time interval $dt$
\begin{equation}
\begin{split}
\hat{U}=\hat{1}+\sum_{i=1}^{k}\left[\dot{\ket{B_i}}\bra{B_i}-\ket{B_i}\bra{\dot{B_i}}\right]dt,
\label{Unitary_lambda_system_generalization}
\end{split}
\end{equation}
We now conclude that the evolution of the dark subspace of the Hamiltonian (\ref{more_general_Hamiltonian}) in the adiabatic regime can be  described with the effective Hamiltonian
\begin{equation}
\hat{H}_{\text{eff}}=\sum_{i=1}^{k}i\left[\dot{\ket{B_i}}\bra{B_i}-\ket{B_i}\bra{\dot{B_i}}\right]=\sum_{i=1}^{k}\hat{H}_i,
\label{Final_effective_Hamiltonian}
\end{equation}
acting in the Hilbert space $\mathcal{H}$. Here $\hat{H}_i$ $(i=1,..,k)$ is a Hamiltonian equivalent to (\ref{effective_hamiltonian}).

\begin{figure}[t]
	\includegraphics[width=0.45\textwidth]{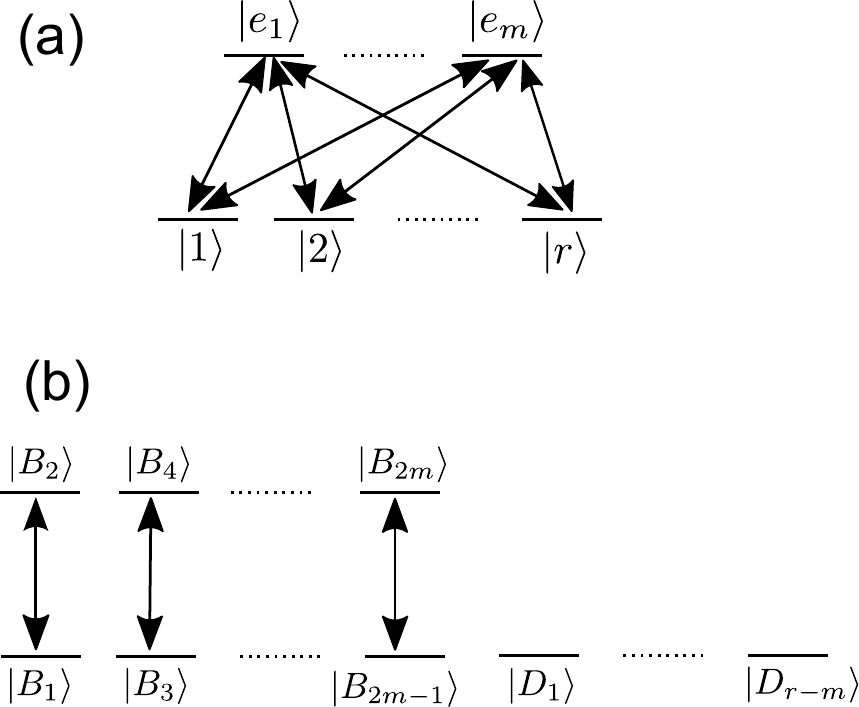}
	\caption{(a) A system having an $r-$fold degenerate ground state and an $m-$fold degenarete excited state. Each arrow denotes the coupling between the corresponding levels. All ground states are coupled to all excited states with the same detuning from the resonance. (b) Under the rotating wave approximation the Morris-Shore transformation  \cite{Morris_Shore} brings this system to at most m driven two-level systems and $r-m$ decoupled dark states (assuming $r\geq m$). The Hamiltonian of this system will then have the form (\ref{more_general_Hamiltonian}). \label{fig:Morris_Shore} }
\end{figure}

Now we will discuss the systems to which our formalism can be applied. One of the simplest cases when the systems' dynamics in the Hilbert space of dimension $n$ is controlled with a Hamiltonian (\ref{more_general_Hamiltonian}), $k<n$, is depicted in Fig \ref{fig:Morris_Shore}.

 Here all the states of $r-$fold degenerate ground space are coupled to all the states of the $m-$fold degenerate excited space with the same excitation frequency. Performing the Morris-Shore transformation \cite{Morris_Shore} and assuming $r\geq m$, the system is brought to at most $m$ coupled pairs. All the other states turn out to be isolated and thus can be associated with the dark states. Going to the frame rotating with the frequency of the excitation and applying the rotating wave approximation, we can describe the system exactly with the Hamiltonian (\ref{more_general_Hamiltonian}) with $n=r+m$ and $k \leq 2m$. Here the bright states $\ket{B_i},\text{ } (1=1,...,k)$ as well as the couplings $g_{i}$ between pairs  will depend on the excitations between the ground and excited states. If one now allows the couplings in the rotating frame to change slowly, so that the adiabatic condition is fulfilled, the states from the dark subspace will evolve according to the effective Hamiltonian ($\ref{Final_effective_Hamiltonian}$). The criterion for the adiabaticity can be formulated as, firstly, the conservation of the number of coupled pairs. In other words, the coupling for any of the pairs never becomes $0$. Secondly, the couplings between the ground and excited states should change much more slowly than the inverse of the smallest coupling strength among the pairs, arising after the Morris-Shore transformation.

 We would like to stress that the Hamiltonian (\ref{more_general_Hamiltonian}) with $k<n$ need not arise necessarily in the system in Fig.~\ref{fig:Morris_Shore}. One may start with the most general case when all $n$  states of the Hilbert space $\mathcal{H}$ are coupled in the rotating frame. In general this Hamiltonian will have no dark states, but if additional conditions are imposed on the couplings, the Hamiltonian may become reducible to the case of the formula (\ref{more_general_Hamiltonian}) with $k<n$. In the case of 
Fig.~\ref{fig:Morris_Shore} this reducibility arises from the abscence of couplings in the excited and ground state manifolds.

\begin{figure}[t]
	\includegraphics[width=0.4\textwidth]{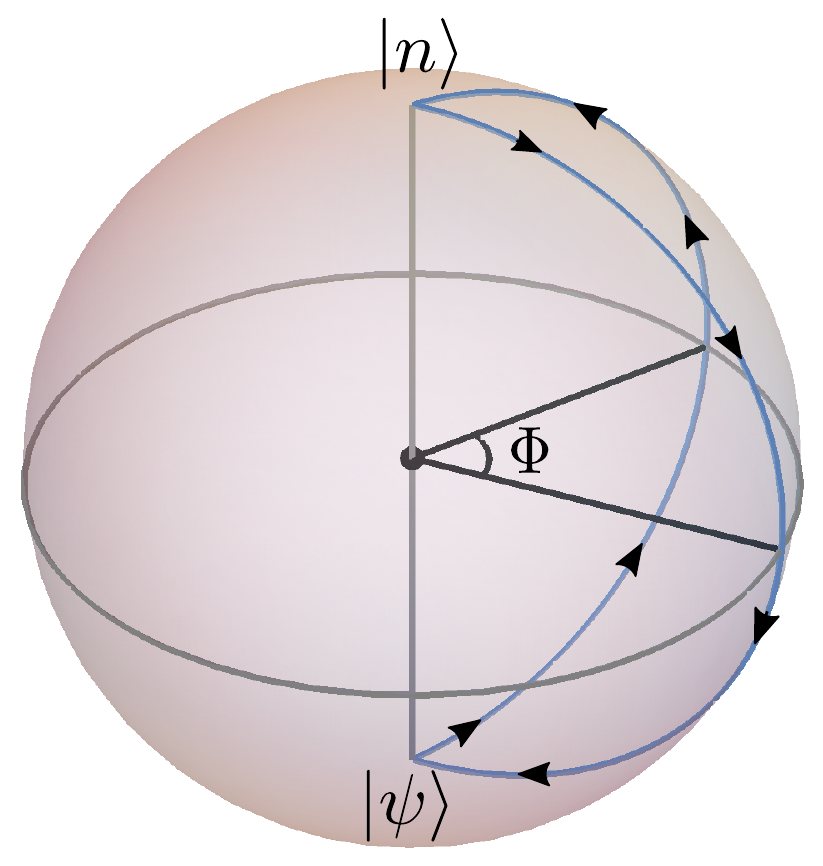}
	\caption{The system initialized in the state $\psi$ moves along a closed path in a two dimensional Hilbert space. At the end of the evolution the system acquires a geometric phase that is equal to $\Phi$, half of the solid angle enclosed by the trajectory. \label{Bloch_sphere} }
\end{figure}

	\textit{Description of quantum gates using effective Hamiltonian.}$\--$Let us now go back to the original Lamda system in Fig \ref{fig:Lambda_system}. We will assume that the system's logical space coincides with the first $n-1$ levels of the ground state space and the system is initialized in this space. We would like to perform a geometric adiabatic gate on the logical space of the system. For that we first switch on only the $n$th coupling $\Omega_n$, so that the bright state initially coincides with the level $\ket{n}$. The logical space is thus the dark space of the Hamiltonian at the beginning of the gate. Let us choose the arbitrary state $\ket{\psi}$ as a linear combination of $\ket{1},\ket{2},.....,\ket{n-1}$ and adiabatically change the couplings in the way that the bright state follows a three-piece trajectory from $t=0$ to $t=t_1$, from $t=t_1$ to $t=t_2$ and from $t=t_2$ to $t=t_3$:
\begin{enumerate}
\item $\begin{aligned}[t]
&\ket{B(t)} = \sin(\theta(t)/2)\ket{\psi}+\cos(\theta(t)/2)\ket{n} \\
&\theta(0)=0,\text{ } \theta(t_1)=\pi
\end{aligned}$
\item $\begin{aligned}[t]
&\ket{B(t)}= e^{i\phi(t)}\ket{\psi} \\
&\phi(t_1)=0,\text{ } \phi(t_2)=\Phi
\end{aligned}$
\item $\begin{aligned}[t]
&\ket{B(t)}= e^{i\Phi}\sin(\theta(t)/2)\ket{\psi}+\cos(\theta(t)/2)\ket{n} \\
& \theta(t_2)=\pi,\text{ } \theta(t_3)=0
\end{aligned}$
\end{enumerate}
Note that the second stage just changes the global phase of the bright state and thus its only meaning is to make the bright state continuous, it does not affect the system that lies in the dark subspace. Therefore this stage can be performed arbitrarily fast without breaking the adibaticity condition. The trajectory of the state of the system is shown in the Fig.~\ref{Bloch_sphere}.

  We can now use the formula (\ref{effective_hamiltonian}) to calculate the effective Hamiltonian for each stage, together with the corresponding unitary $\hat{T}e^{-i\int_{\text{start}}^{\text{end}} \hat{H}_{\text{eff}}(t) dt}$ and obtain
\begin{enumerate}
	\item $\begin{aligned}[t]
	&\hat{H}_{\text{eff}}=i\frac{\dot{\theta}}{2}(\ket{\psi}\bra{n}-\ket{n}\bra{\psi}) \\
	&\hat{U}_1=e^{\frac{\pi}{2}(\ket{\psi}\bra{n}-\ket{n}\bra{\psi})}=\ket{\psi}\bra{n}-\ket{n}\bra{\psi}
	\end{aligned}$
	\item $\begin{aligned}[t]
	&\hat{H}_{\text{eff}}=-2\dot{\phi}\ket{\psi}\bra{\psi} \\
	&\hat{U}_2=e^{2i\Phi\ket{\psi}\bra{\psi}}=\hat{1}-\ket{\psi}\bra{\psi}+e^{2i\Phi}\ket{\psi}\bra{\psi}
	\end{aligned}$
	\item $\begin{aligned}[t]
	\hat{H}_{\text{eff}}&=\frac{\dot{\theta}}{2}(i \cos(\Phi)(\ket{\psi}\bra{n}-\ket{n}\bra{\psi})\\
	&-\sin(\Phi)(\ket{\psi}\bra{n}+\ket{n}\bra{\psi}) \\
	\hat{U}_3&=-\cos(\Phi)(\ket{\psi}\bra{n}-\ket{n}\bra{\psi})\\
	&-i\sin(\Phi)(\ket{\psi}\bra{n}+\ket{n}\bra{\psi})
	\end{aligned}$
\end{enumerate}
We note that after the third stage the bright state returns back to $\ket{n}$, which indicates that the dark subspace at the end of the gate coincides with the logical space. Combining the action of the three stages we obtain
\begin{equation}
\hat{U}=\hat{U}_3\hat{U}_2\hat{U}_1=e^{-i \Phi}\ket{\psi}\bra{\psi}+e^{i \Phi}\ket{n}\bra{n}
\label{Gate}
\end{equation}
The action on the state $\ket{n}$ is irrelevant as the effective Hamiltonian only describes the evolution of the vectors from the dark subspace correctly. The state $\ket{\psi}$ on the other hand obtains a phase factor $-\Phi$. This phase factor is equal to half of the solid angle the trajectory of the state $\ket{\psi}$ traces on the Bloch sphere in Fig.~\ref{Bloch_sphere}, that coincides with the classical result \cite{Berry,Shapere_Wilczek}.

If $n=3$ the gate above is a single qubit gate on the levels $\ket{1},\ket{2}$. If one puts these two levels on the other Bloch sphere, this gate can be thought of as rotation by the angle $\Phi$ around the axis on this sphere, which is defined by the choice of $\ket{\psi}$. If $n=5$ the gate above is a two qubit gate, as the dimension of the logical space is four. If also $\ket{\psi}=\ket{4}$, this is a CPHASE gate, which in combination with the universal single qubit gates gives a complete set of gates in the space of two qubits. We also note that if one does not restrict oneself to closed trajectories, the effective Hamiltonian describes the STIRAP process.

\textit{Discussion.}$\--$We have shown that the adiabatic dynamics of a state from the dark subspace of the Hamiltonian (\ref{more_general_Hamiltonian}) can be described with an effective Hamiltonian (\ref{Final_effective_Hamiltonian}). Due to the presence of derivatives in this Hamiltonian, the equation of motion for the wave function turns out to be invariant with respect to the reparametrization of time $\tau=f(t)$, which reflects the geometric nature of the evolution.
We stress that the calculation of the effective Hamiltonian (\ref{Final_effective_Hamiltonian}) requires that one brings the initial dynamical Hamiltonian to the form (\ref{more_general_Hamiltonian}), which involves the identification of the instantaneous orthonormal bright states of the system. But for this, it is not necessary to compute the basis in the dark subspace. For large dimensionalities of the Hilbert space one often encounters a situation, when the number of dark states is much greater than the number of bright states. If the Hamiltonian (\ref{more_general_Hamiltonian}) takes the simplified form of (\ref{Lambda_Hamiltonian}) with two bright states and one of them constant, it is possible to easily parametrize all the dark states using the coordinates on the sphere \cite{Recati_Calarco_Zanardi_Cirac_Zoller}. But if the situation is more complicated with two or more bright states changing in time, one can no longer easily parametrize the dark subspace. This would involve the Gram-Schmidt orthogonalization procedure which is a recursive process that can take a long time even with the use of powerful computers. In that case the effective Hamiltonian will be very useful because it allows to avoid this procedure.

In this paper we do not discuss the fidelities of the quantum gates obtained through the geometric evolution, but instead give an alternative description of the latter. The effects of noise on the adiabatic evolution were extensively studied theoretically \cite{Albash},  recent experimental results show evidence for a selectivity of noise, which the system is sensitive to \cite{Yale_Awschalom}. A description of noise within our effective Hamiltonian formalism is out of scope of this paper and will be adressed in future work.

%%%%%%%%%%%%%%%%%%%%%%%%%%%%%%%%%%%%%%%%%%%%%%%%%%%%%%%%%%%%%%%
%%% Acknowledgement
%%%%%%%%%%%%%%%%%%%%%%%%%%%%%%%%%%%%%%%%%%%%%%%%%%%%%%%%%%%%%%%
\textit{Acknowledgements.}$\--$We thank Maximillian Russ for helpful discussions. We acknowledge funding from DFG through SFB767.

%%%%%%%%%%%%%%%%%%%%%%%%%%%%%%%%%%%%%%%%%%%%%%%%%%%%%%%%%%%%%%%
%%% Appendix
%%%%%%%%%%%%%%%%%%%%%%%%%%%%%%%%%%%%%%%%%%%%%%%%%%%%%%%%%%%%%%%

\clearpage
\onecolumngrid
\appendix

\section{Effective Hamiltonian for a generalized Lambda system} 
\label{appA}

In this section, we derive Eqs.~(\ref{Unitary_lambda_system}) and (\ref{effective_hamiltonian}) of the main text.
We consider a generalized Lamda system with $n+1$ levels, for which the first $n$ levels, forming a Hilbert space $\mathcal{H}$ are resonantly coupled to the remaining level (Fig.~\ref{fig:Lambda_system}). The Hamiltonian of such a system in the rotating frame is
\begin{equation}
\hat{H}=\Omega\sum_{i=1}^{n}r_i\left(e^{i\phi_i}\ket{i}\bra{e}+e^{-i\phi_i}\ket{e}\bra{i}\right).
\label{Aone_bright_state_hamiltonian_init}
\end{equation}
By introducing the bright state $\ket{B}=\sum_{i=1}^{n}r_ie^{i\phi_i}\ket{i}$ the Hamiltonian may be rewritten as
\begin{equation}
\hat{H}=\Omega\left(\ket{B}\bra{e}+\ket{e}\bra{B}\right),\\
\label{Aone_bright_state_hamiltonian}
\end{equation}

%\begin{figure}[h]
%	\includegraphics[width=0.45\textwidth]{Lambda_system}
%	\caption{Energy level diagram of the generalized Lambda-system. The lower $n$ states $\ket{i}$ ($i=1,...,n$) are coupled to the excited state $\ket{e}$. The system %has the instantaneous $(n-1)-$dimensional dark subspace, in which it resides in accordance with the adiabatic theorem.
%		\label{fig:Lambda_system}}
%\end{figure}

The $n-1$ states spanning the Hilbert space $\mathcal{H}$ are eigenvectors of this Hamiltonian with zero eigenvalue, thus forming the dark subspace of this Hamiltonian. We assume that Hamiltonian (\ref{Aone_bright_state_hamiltonian}) is time dependent, such that the dark subspace also depends on time and is defined with a time dependent orthonormal basis $\ket{\psi_1(t)}$, $\ket{\psi_2(t)}$,......, $\ket{\psi_{n-1}(t)}$. We assume the system to be initialized in an instantaneous dark state of its Hamiltonian and to subsequently evolve in the adiabatic regime, so at any time $t$ the state of the system can be expressed as $\ket{\psi_s}=\sum_{i=1}^{n-1}c_i(t)\ket{\psi_i(t)}$. 

 The most general state the system may be in is $\ket{\psi}=\sum_{i=1}^{n-1}c_i(t)\ket{\psi_i(t)}+p\ket{B}$. According to the Schr\"{o}dinger equation 
\begin{equation}
	i\hbar\frac{\partial}{\partial t}\left(\sum_{i=1}^{n-1}c_i(t)\ket{\psi_i(t)}+p\ket{B}\right)=p\hat{H}\ket{B},
\end{equation}
because the Hamiltonian acts trivially on the dark states. In other words, one may write
\begin{equation}
\sum_{j=1}^{n-1}\left(\dot{c}_j(t)\ket{\psi_j(t)}+c_j(t)\ket{\dot{\psi}_j(t)}\right)+\dot{p}\ket{B}+p\ket{\dot{B}}=p\hat{H}\ket{B}.
\end{equation}
Multiplying this equation by $\bra{\psi_i}$ from the left we obtain
\begin{equation}
\dot{c}_i(t)=-\sum_{j=1}^{n-1}c_j(t)\braket{\psi_i(t)|\dot{\psi}_j(t)}-p\braket{\psi_i|\dot{B}}.
\end{equation}
In the adiabatic limit $p\rightarrow 0$, we find
\begin{equation}
\dot{c}_i(t)=-\sum_{j=1}^{n-1}c_j(t)\braket{\psi_i(t)|\dot{\psi}_j(t)}.
\label{Ac_coefficients}
\end{equation}

Let us now concentrate on the evolution of the dark subspace during an infinitesimal time interval $dt$. A general state of the system in the adiabatic limit $\ket{\psi_s}=\sum_{i=1}^{n-1}c_i(t)\ket{\psi_i(t)}$ evolves into
\begin{eqnarray}
\begin{split}
\ket{\psi_s}=&\sum_{i=1}^{n-1}c_i(t)\ket{\psi_i(t)}\rightarrow \sum_{i=1}^{n-1}c_i(t+dt)\ket{\psi_i(t+dt)}=\\
=&\sum_{i=1}^{n-1}(c_i(t)+\dot{c}_i(t)dt)(\ket{\psi_i(t)}+\ket{\dot{\psi}_i(t)}dt),
\end{split}
\end{eqnarray} 
which using Eq. (\ref{Ac_coefficients}) up to the linear terms in $dt$ becomes
\begin{equation}
\sum_{i=1}^{n-1}
\left(c_i(t)\ket{\psi_i(t)}+c_i(t)\ket{\dot{\psi}_i(t)}dt\right)
-\sum_{i,j=1}^{n-1}c_j(t)\braket{\psi_i(t)|\dot{\psi}_j(t)}\ket{\psi_i(t)}dt.
\end{equation}
Introducing the operator 
\begin{equation}
\begin{split}
	\hat{O}_D(dt)&=\sum_{i=1}^{n-1}\ket{\psi_i(t)}\bra{\psi_i(t)}+dt\sum_{i=1}^{n-1}\ket{\dot{\psi}_i(t)}\bra{\psi_i(t)}\\
	&-dt\sum_{i,j=1}^{n-1}\braket{\psi_i(t)|\dot{\psi}_j(t)}\ket{\psi_i(t)}\bra{\psi_j(t)},\\
\end{split}
\end{equation}
 it follows that 
\begin{equation}
\ket{\psi_s}\rightarrow \hat{O}_D(dt)\ket{\psi_s}.
\label{Atransformation}
\end{equation}

 Here the dark states form an $(n-1)-$dimensional subspace of the $n-$dimensional Hilbert space $\mathcal{H}$. Therefore $\hat{O}_D(dt)$ can be viewed as an operator acting in the space $\mathcal{H}$. Introducing the projector onto the dark space
\begin{equation}
     \hat{P}_D=\sum_i\ket{\psi_i(t)}\bra{\psi_i(t)},
     \label{Adark_space_projector}
\end{equation}
we arrive at 
\begin{equation}
\hat{O}_D=\hat{P}_D+\dot{\hat{P}}_D\hat{P}_Ddt.
\label{Adark_subspace_transformer}
\end{equation}
The expression for $\hat{O}_D$ can be even further simplified, if one uses the fact that $\hat{P}_D=\hat{1}-\hat{P_B}$, where $\hat{1}$ is the identity operator acting in the Hilbert space $\mathcal{H}$ and $\hat{P_B}=\ket{B}\bra{B}$. We obtain
\begin{equation}
\begin{split}
\hat{O}_D&=\hat{1}-\hat{P}_B-\dot{\hat{P}}_B(\hat{1}-\hat{P}_B)dt=\\
&=\hat{1}-\hat{P}_B+\left[\braket{\dot{B}|B}\ket{B}\bra{B}-\ket{B}\bra{\dot{B}}\right]dt.
\label{Adark_subspace_transformer_explicit}
\end{split}
\end{equation}
This operator transforms the basis vectors of the dark subspace and because it arose from the Schr\"{o}dinger equation with the Hamiltonian (\ref{Aone_bright_state_hamiltonian}) in the adiabatic limit, we can conclude that the orthonormal basis vectors from the dark subspace, corresponding to time $t$, are transformed into orthonormal vectors from the dark subspace, corresponding to time $t+dt$. $\hat{O}_D$ is not unitary, as it takes the bright state to 0, whereas we can make $\hat{O}_D$ unitary if we complement it with an operator performing the following transformation 
\begin{equation}
e^{i\alpha(t)}\ket{B(t)}\rightarrow e^{i\alpha(t+dt)}\ket{B(t+dt)}.
\end{equation}

In analogy with the dark subspace, the operator performing this transformation is
\begin{equation}
\hat{O}_B=\hat{P}_B+\left[i\dot{\alpha}(t)\ket{B}\bra{B}+\dot{\ket{B}}\bra{B}\right]dt,
\end{equation}
Now we can define the unitary transformation $\hat{U}$, acting in the whole Hilbert space $\mathcal{H}$, such that it yields the correct evolution of the vectors in the dark subspace
\begin{equation}
\begin{split}
\hat{U}=\hat{O}_D+\hat{O}_B=\hat{1}&+\left[\dot{\ket{B}}\bra{B}-\ket{B}\bra{\dot{B}}\right]dt+\\
&+\left(i\dot{\alpha}(t)+\braket{\dot{B}|B}\right)\ket{B}\bra{B}dt.
\label{AUnitary_lambda_system}
\end{split}
\end{equation}
The operator $\hat{U}$ is not uniquely defined as there is still some freedom left in defining $\dot{\alpha}(t)$. From the normalization condition $\braket{B|B}=1$ it follows that $\braket{\dot{B}|B}$ is purely imaginary; furthermore we can define $\dot{\alpha}(t)$ such that $i\dot{\alpha}(t)+\braket{\dot{B}|B}=0$.
We then obtain the unitary
\begin{eqnarray}
\hat{U}=\hat{1}+\left[\dot{\ket{B}}\bra{B}-\ket{B}\bra{\dot{B}}\right]dt,
\end{eqnarray}
which acts in the whole space $\mathcal{H}$ and generates the correct evolution in the dark subspace. Using the relation $\hat{U}=e^{-i\hat{H}dt}=\hat{1}-i\hat{H}dt$, we can view the evolution of the state in the dark subspace in the adiabatic regime as if a time dependent effective Hamiltonian
\begin{equation}
\hat{H}_{\text{eff}}=i\left[\dot{\ket{B}}\bra{B}-\ket{B}\bra{\dot{B}}\right]
\label{Aeffective_hamiltonian}
\end{equation}
was acting in the Hilbert space $\mathcal{H}$. From here on, we assume $\hbar=1$.

\section{Effective Hamiltonian for a more general quantum system with a dark space}
\label{appB}

In this section, we derive Eqs.~(\ref{Unitary_lambda_system_generalization}) and (\ref{Final_effective_Hamiltonian}) of the main text.
Let us now consider a more general Hamiltonian of the form
\begin{equation}
\hat{H}=\sum_{i,j=1}^{k}g_{ij}\ket{B_i}\Bra{B_j}+g_{ij}^*\ket{B_j}\Bra{B_i},
\label{Amore_general_Hamiltonian}
\end{equation}
acting in some Hilbert space $\mathcal{H}$ of dimension $n$, where $\ket{B_i}$ are time dependent states in $\mathcal{H}$, forming an orthonormal set of vectors at any instant in time. Note that the Hamiltonian (\ref{Aone_bright_state_hamiltonian}) is a special case of (\ref{Amore_general_Hamiltonian}), with two bright states, one of them being time-independent. We point out that the Hamiltonian (\ref{Amore_general_Hamiltonian}) could always be brought to diagonal form with appropriately chosen bright states, but for our purposes it is not necessary to assume this.

From now on we will assume that the eigenstates of the Hamiltonian (\ref{Amore_general_Hamiltonian}) in the subspace spanned by the vectors $\ket{B_{i}}$ have nonzero instantaneous eigenvalues and that the adiabatic condition with respect to these eigenvalues is fulfilled. Thus, if the system starts in the instantaneous dark subspace of this Hamiltonian, it remains in the dark subspace, in accordance with the adiabatic theorem. Equations (\ref{Atransformation}), (\ref{Adark_space_projector}) and (\ref{Adark_subspace_transformer}) remain unchanged with the number of bright states increasing, while equation (\ref{Adark_subspace_transformer_explicit}) is replaced by
\begin{equation}
\begin{split}
\hat{O_D}&=\hat{1}-\sum_{i=1}^{k}\hat{P}_{B_i}-\left(\sum_{i=1}^{k}\dot{\hat{P}}_{B_i}\right)\left(\hat{1}-\sum_{i=1}^{k}\hat{P}_{B_i}\right)dt\\
&=\hat{1}-\sum_{i=1}^{k}\hat{P}_{B_i}+\left[\sum_{i,j=1}^{k}\braket{\dot{B}_i|B_j}\ket{B_i}\bra{B_j}-\sum_{i=1}^{k}\ket{B_i}\bra{\dot{B}_i}\right]dt.
\end{split}
\end{equation}
This operator is again non-unitary in exactly the same sense as the operator in equation (\ref{Adark_subspace_transformer_explicit}). To make $\hat{O}_D$ unitary we can complement it with an operator, performing in general the following transformation
\begin{equation}
\ket{B_i(t)}\rightarrow \sum_{j=1}^{k}\tilde{U}_{ij}(dt)\ket{B_j(t+dt)},
\label{Abright_states_transition}
\end{equation}
where $\tilde{U}_{ij}(dt)$ is an arbitrary infinitesimal unitary transformation in the subspace spanned by the vectors $\ket{B_j(t+dt)}$. If we introduce a general Hermitian matrix $A_{ij}$, we can write
\begin{equation}
\tilde{U}_{ij}(dt)=e^{iA_{ij}dt}\simeq\delta_{ij}+iA_{ij}dt
\end{equation}
and thus equation (\ref{Abright_states_transition}) takes the form
\begin{equation}
\ket{B_i(t)}\rightarrow \ket{B_i(t)}+\ket{\dot{B}_i(t)}dt+i\sum_{j=1}^{k}A_{ij}\ket{B_j}dt.
\end{equation}
The operator performing this transformation is 
\begin{equation}
\hat{O}_B=\sum_{i=1}^{k}\hat{P}_{B_i}+\sum_{i=1}^{k}\ket{\dot{B}_i(t)}\bra{B_i(t)}dt+i\sum_{i,j=1}^{k}A_{ij}\ket{B_j(t)}\bra{B_i(t)}dt.
\end{equation}
In full analogy to (\ref{AUnitary_lambda_system}), we can build the unitary transformation acting in the Hilbert space $\mathcal{H}$ by adding $O_B$ and $O_D$
\begin{equation}
\begin{split}
\hat{U}=\hat{O}_D+\hat{O}_B=\hat{1}&+\sum_{i=1}^{k}\left[\dot{\ket{B_i}}\bra{B_i}-\ket{B_i}\bra{\dot{B_i}}\right]dt+\\
&+\sum_{i,j=1}^{k}(iA_{ji}(t)+\braket{\dot{B_i}|B_j})\ket{B_i}\bra{B_j}dt.
\label{AUnitary_lambda_system_new}
\end{split}
\end{equation}
The arbitrariness in $A_{ij}(t)$ can be removed if we choose $A_{ji}(t)=i\braket{\dot{B_i}|B_j}$. Note that this definition is consistent with the Hermitian property of A, as  
\begin{equation}
i\braket{\dot{B_i}|B_j}=-i\braket{B_i|\dot{B}_j}=-i\braket{\dot{B}_j|B_i}^*=(i\braket{\dot{B}_j|B_i})^*.
\end{equation}

For the unitary operator $\hat{U}$ we then obtain
\begin{equation}
\hat{U}=\hat{1}+\sum_{i=1}^{k}\left[\dot{\ket{B_i}}\bra{B_i}-\ket{B_i}\bra{\dot{B_i}}\right]dt.
\end{equation}

We now conclude that the evolution of the dark subspace of the Hamiltonian (\ref{Amore_general_Hamiltonian}), acting in the Hilbert space $\mathcal{H}$ can be  described with the effective Hamiltonian
\begin{equation}
\hat{H}_{\text{eff}}=\sum_{i=1}^{k}i\left[\dot{\ket{B_i}}\bra{B_i}-\ket{B_i}\bra{\dot{B_i}}\right]=\sum_{i=1}^{k}\hat{H}_i,
\label{AFinal_effective_Hamiltonian}
\end{equation}
acting in the Hilbert space $\mathcal{H}$. Here $\hat{H}_i$ $(i=1,..,k)$ is a Hamiltonian equivalent to (\ref{Aeffective_hamiltonian}).

\section{Comparison of Non-Abelian Berry connection to the effective Hamiltonian}
\label{appC}
In this section we will consider the system shown in figure (\ref{fig:Lambda_system}) with three ground states ($n=3$) and show that the universal set of single-qubit gates on two of them can be equivalently described either with the language of non-Abelian Berry connection or with the effective Hamiltonian.

The dark states of the Hamiltonian (\ref{Hamiltonian}) can be parametrized with the angles of the sphere $\theta_1$, $\theta_2$, if one parametrizes coupling coefficients $r_i$ as \cite{Recati_Calarco_Zanardi_Cirac_Zoller}
\begin{eqnarray}
\begin{split}
r_1&=\sin(\theta_1),\\
r_2&=\cos(\theta_1)\sin(\theta_2),\\
r_3&=\cos(\theta_1)\cos(\theta_2).\\
\end{split}
\label{couplings_vs_parameters}
\end{eqnarray}
For the dark states one then obtains \cite{Recati_Calarco_Zanardi_Cirac_Zoller}
\begin{eqnarray}
\begin{split}
\ket{d_1}&=\cos(\theta_1)\ket{1}-\sin(\theta_1)(e^{i\phi_2}\sin(\theta_2)\ket{2}+e^{i\phi_3}\cos(\theta_2)\ket{3}),\\
\ket{d_2}&=e^{i\phi_2}\cos(\theta_2)\ket{2}-e^{i\phi_3}\sin(\theta_2)\ket{3}.\\
\label{dark_states_parameters}
\end{split}
\end{eqnarray}
Treating $\theta_1,\theta_2,\phi_2,\phi_3$ as parameters ($\lambda_k$, $k=\{1,2,3,4\}$) and calculating the Berry connection $A_{k}=\braket{d_i|\frac{\partial}{\partial\lambda_k}|d_j}$ one obtains
\begin{eqnarray}
\begin{split}
A_{\theta_1}&=\left(\begin{array}{c c}
0     &  0 \\
0     &  0 
\end{array}\right),\\
A_{\theta_2}&=\left(\begin{array}{c c}
0     &  \sin(\theta_1) \\
-\sin(\theta_1)     &  0 
\end{array}\right),\\
A_{\phi_2}&=i\left(\begin{array}{c c}
\sin^2(\theta_1)\sin^2(\theta_2)     &  \sin(\theta_1)\sin(\theta_2)\cos(\theta_2) \\
\sin(\theta_1)\sin(\theta_2)\cos(\theta_2)     &  \cos^2(\theta_2) 
\end{array}\right),\\
A_{\phi_3}&=i\left(\begin{array}{c c}
\sin^2(\theta_1)\cos^2(\theta_2)     &  -\sin(\theta_1)\sin(\theta_2)\cos(\theta_2) \\
-\sin(\theta_1)\sin(\theta_2)\cos(\theta_2)     &  \sin^2(\theta_2) 
\end{array}\right).\\
\end{split}
\end{eqnarray}
We assume a loop that starts with the parameters $\theta_1=\theta_2=\phi_2=\phi_3=0$, such that the dark subspace is spanned by $\{\ket{1},\ket{2}\}$. If we perform a closed loop in parameter space, forcing the dark subspace to undergo a closed loop in the Hilbert space, a unitary on the dark subspace will be induced, that in the basis $\{\ket{1},\ket{2}\}$ takes the form
\begin{equation}
U=\hat{P}\exp\left(-\oint\sum_{k=1}^4A_kd\lambda_k\right),
\label{unitary_Berry_connection}
\end{equation}
where $\hat{P}$ corresponds to the operation of path ordering. 

If we only vary the two parameters $\theta_1$ and $\theta_2$, Eq. (\ref{unitary_Berry_connection}) takes the form
\begin{equation}
U_y=\exp\left(-i\sigma_y\oint\sin(\theta_1)d\theta_2\right),
\label{unitary_sigmaY}
\end{equation} 
where $\sigma_y$ is the Pauli matrix.

If we only vary $\theta_2$ and $\phi_3$, Eq. (\ref{unitary_Berry_connection}) takes the form
\begin{equation}
U_z=\exp\left(-i\oint\left(\begin{array}{c c}
0    & 0 \\
0    &  \sin^2(\theta_2) 
\end{array}\right)d\phi_3\right)
\label{unitary_Z}.
\end{equation} 

These two types of loops generate rotations around Y and Z axes respectively. Thus, the two operations do not commute and are sufficient to generate a universal set of gates. 

We can alternatively analyze these loops using the effective Hamiltonian, equation (\ref{laser_couplings_effective Hamiltonian}). Given the loop in the parameter space $(\theta_1(t),\theta_2(t), t=[0,T])$, using the equations (\ref{laser_couplings_effective Hamiltonian}),    (\ref{couplings_vs_parameters}) one can construct the effective Hamiltonian
\begin{eqnarray}
\begin{split}
\hat{H}_{\text{eff}}=&i\sum_{i,j}(r_j\dot{r}_i-r_i\dot{r}_j)\ket{i}\bra{j}\\
=&i(\sin(\theta_2)\dot{\theta}_1-\sin(\theta_1)\cos(\theta_1)\cos(\theta_2)\dot{\theta}_2)(\ket{1}\bra{2}-\ket{2}\bra{1})\\
&+i(\cos(\theta_2)\dot{\theta}_1+\sin(\theta_1)\cos(\theta_1)\sin(\theta_2)\dot{\theta}_2)(\ket{1}\bra{3}-\ket{3}\bra{1})\\
&+i\cos^2(\theta_1)\dot{\theta}_2(\ket{2}\bra{3}-\ket{3}\bra{2}).\\
\end{split}
\label{effective Hamiltonian for a loop}
\end{eqnarray}

We numerically solved the Schr\"{o}dinger equation to obtain the final unitary 
\begin{equation}
U=\hat{T}\exp\left(-\int_0^T\hat{H}_{\text{eff}}(t)dt\right).
\label{unitary_effective_Hamiltonian}
\end{equation}
Restricted to the space of $\{\ket{1},\ket{2}\}$, the result exactly coincides with the unitaries obtained with Eqs. (\ref{unitary_sigmaY}), (\ref{unitary_Z}) obtained for the Berry connection. We also did the same check for the $U_z$ gates with the same results. 

We note that although in our example the Berry connection approach did not require integration and thus is easier to implement, it relies on the explicit parametrization of the dark states, Eq. (\ref{dark_states_parameters}). In contrast, the effective Hamiltonian (\ref{laser_couplings_effective Hamiltonian}) does not require explicit parametrization of the coupling parameters and thus could be used without it, given only the dependence of the couplings on time. This will still hold in more general cases, when the number of bright states is larger than one and the dark subspace cannot be parametrized so easily. Then the effective Hamiltonian would allow to calculate the unitary arising from purely geometric evolution with much lower numerical cost, without the necessity to numerically orthogonalize the dark subspace.
%%%%%%%%%%%%%%%%%%%%%%%%%%%%%%%%%%%%%%%%%%%%%%%%%%%%%%%%%%%%%%%
%%% Bibliography
%%%%%%%%%%%%%%%%%%%%%%%%%%%%%%%%%%%%%%%%%%%%%%%%%%%%%%%%%%%%%%%


\begin{thebibliography}{99}
	
	\bibitem{Preskill}
	J. Preskill,
	\textit{Quantum Computing in the NISQ era and beyond},
	arXiv:1801.00862.

    \bibitem{Campbell}
     E. T. Campbell, B. M. Terhal, and C. Vuillot, 
     \textit{Roads towards fault-tolerant universal quantum computation},
     Nature {\bf 549}, 172 (2017).

	\bibitem{Berry}
    M. V. Berry,
	\textit{Quantal phase factors accompanying adiabatic changes},
	Proceedings of the royal society A {\bf 392}, 1802  (1984).
	
	\bibitem{Zanardi_Rosetti}
	P. Zanardi, M. Rasetti,
	\textit{Holonomic quantum computation},
	Physics Letters A {\bf 264}, 94  (1999).
	
	\bibitem{Sjoeqvist3}
	Erik Sjöqvist,
	\textit{Geometric phases in quantum information},
	International journal of Quantum Chemistry 115, 1311  (2015).
	
	\bibitem{Shi-Liang Zhu_Paolo Zanardi}
	S.-L. Zhu and P. Zanardi,
	\textit{Geometric quantum gates that are robust against stochastic control errors},
	Phys. Rev. A {\bf 72}, 020301 (2005).
	
	\bibitem{Childs_Edward Farhi_John Preskill}
	A. M. Childs, E. Farhi, J. Preskill,
	\textit{Robustness of adiabatic quantum computation},
	Phys. Rev. A {\bf 65}, 012322 (2001).
	
	\bibitem{NMR_Jones_Vedral_Ekert_Giuseppe Castagnoli}
	J. A. Jones, V. Vedral, A. Ekert
	and G. Castagnoli,
	\textit{Geometric quantum computation
		using nuclear magnetic resonance},
	Nature {\bf 403}, 869 (2000).
	
	
	\bibitem{Duan_Cirac_Zoller}
	L.-M. Duan, J. I. Cirac, P. Zoller,
	\textit{Geometric Manipulation of Trapped Ions for Quantum Computation},
	Science {\bf 292}, 1695 (2001).
	
	\bibitem{Kis_Renzoni}
	Z. Kis and F. Renzoni,
	\textit{Qubit rotation by stimulated Raman adiabatic passage},
	Phys. Rev. A {\bf 65}, 032318 (2002).
	
	\bibitem{Moeller_Madsen_Moelmer}
	D. Møller, L.B. Madsen, and K. Mølmer,
	\textit{Geometric phase gates based on stimulated Raman adiabatic passage in tripod systems},
	Phys. Rev. A {\bf 75}, 062302 (2007).
	
	\bibitem{Trapped_ions_Toyoda_Uchida_Noguchi_Haze_Urabe}
	K. Toyoda, K. Uchida, A. Noguchi, S. Haze, and S. Urabe,
	\textit{Realization of holonomic single-qubit operations},
	Phys. Rev. A {\bf 87}, 052307 (2013).
	
	\bibitem{Kamleitner_Solinas_Mueller_Shnirman_Moettoenen}
	I. Kamleitner, P. Solinas, C. Müller, A. Shnirman, and M. Möttönen,
	\textit{Geometric quantum gates with superconducting qubits},
	Phys. Rev. B {\bf 83}, 214518 (2011).
	
	\bibitem{Sjoeqvist}
	E. Sj\"oqvist, D. M. Tong, L. M. Andersson, B. Hessmo, M. Johansson and K. Singh,
	\textit{Non-adiabatic holonomic quantum computation},
	New Journal of Physics {\bf 14}  (2012).
	
	\bibitem{Sjoeqvist2}
	E. Sj\"oqvist, V. A. Mousolou, C. M. Canali,
	\textit{Conceptual aspects of geometric quantum computation},
	Quantum Inf. Process. {\bf 15}, 3995 (2016).
	
	\bibitem{Abdumalikov}
	A. A. Abdumalikov Jr, J. M. Fink, K. Juliusson, M. Pechal, S. Berger, A. Wallraff, S. Filipp,
	\textit{Experimental realization of non-Abelian non-adiabatic geometric gates},
	Nature {\bf 496}, 482 (2013).
		
	\bibitem{Feng}
	G. Feng, G. Xu and G. Long,
	\textit{Experimental Realization of Nonadiabatic Holonomic Quantum Computation},
	Phys. Rev. Lett. {\bf 110}, 190501 (2013). 
		
	\bibitem{Balasubramanian}
	S. Arroyo-Camejo, A. Lazariev, S.W. Hell, G. Balasubramanian,
	\textit{Room temperature high-fidelity holonomic single-qubit gate on a solid-state spin},
	Nature Communications {\bf 5}, 4870 (2014).
		
	\bibitem{Zu}
	C. Zu, W.-B. Wang, L. He, W.-G. Zhang, C.-Y. Dai, F. Wang, L.-M. Duan,
	\textit{Experimental realization of universal geometric quantum gates with solid-state spins},
	Nature {\bf 514}, 72 (2014).
		
	\bibitem{Zhou}
	B.B. Zhou, P.C. Jerger, V.O. Shkolnikov, F. J. Heremans, G. Burkard, D. D. Awschalom,
	\textit{Holonomic Quantum Control by Coherent Optical Excitation in    Diamond},
	Phys. Rev. Lett. {\bf 119}, 140503 (2017). 
	
	\bibitem{Shapere_Wilczek}
	A. Shapere, F. Wilczek,
	\textit{Geometric phases in physics},
	World Scientific (1989).
	
	\bibitem{Wilczek_Zee}
	F. Wilczek and A. Zee,
	\textit{Appearance of Gauge Structure in Simple Dynamical Systems},
	Phys. Rev. Lett. {\bf 52}, 2111  (1984).
	
	\bibitem{Recati_Calarco_Zanardi_Cirac_Zoller}
    A. Recati, T. Calarco, P. Zanardi, J. I. Cirac, and P. Zoller,
	\textit{Holonomic quantum computation with neutral atoms},
	Phys. Rev. A {\bf 66}, 032309  (2002).
	
	\bibitem{Born_Fock}
    M. Born and V. A. Fock,
	\textit{Beweis des Adiabatensatzes},
	Zeitschrift für Physik A {\bf 51}, 165  (1928).
		
	\bibitem{Fleischhauer}
	M. Fleischhauer, A. Imamoglu, and J. P. Marangos,
	\textit{Electromagnetically induced transparency: Optics in coherent media},
	Rev. Mod. Phys. {\bf77}, 633 (2005).
	
	\bibitem{Morris_Shore}
	J. R. Morris and B. W. Shore,
	\textit{Reduction of degenerate two-level excitation to independent two-state systems},
	Phys. Rev. A {\bf 27}, 906  (1983).
	
	\bibitem{Albash}
	T. Albash, S. Boixo, D. A. Lidar, and Paolo Zanardi,
	\textit{Quantum adiabatic Markovian master equations},
	New J. Phys. {\bf 17}, 129501  (2015).
	
	\bibitem{Yale_Awschalom}
	C. G. Yale, F. J. Heremans, B. B. Zhou, A. Auer, G. Burkard and D. D. Awschalom,
	\textit{Optical manipulation of the Berry phase	in a solid-state spin qubit},
	Nature Photonics {\bf10}, 184–189 (2016).
\end{thebibliography}
\end{document}